\documentclass[proceedings]{JHEP3}

\PrHEP{PrHEP hep2001}                   
\conference{International Europhysics Conference on HEP}                

\usepackage{epsfig}                   

\newcommand{\be}{\begin{eqnarray}}
\newcommand{\ee}{\end{eqnarray}}
\newcommand{\nee}{\nonumber\end{eqnarray}}

\def\lsim{\raise0.3ex\hbox{$\;<$\kern-0.75em\raise-1.1ex\hbox{$\sim\;$}}}
\def\gsim{\raise0.3ex\hbox{$\;>$\kern-0.75em\raise-1.1ex\hbox{$\sim\;$}}}

\title{General Flavour Blind MSSM and CP Violation}

\author{\speaker{Alfred Bartl} \\                          
        Institut f\"ur Theoretische Physik, 
  Universit\"at Wien, A-1090, Vienna, Austria \\           
        E-mail: \email{bartl@ap.univie.ac.at}}                     

\author{Thomas Gajdosik\\                                                  
        University of Alabama, Tuscaloosa, Alabama 35487, USA\\                                    
        E-mail: \email{garfield@qhepu3.oeaw.ac.at}}
\author{Enrico Lunghi\\
        Deutches Elektronen Synchrotron DESY, Hamburg\\
        E-mail: \email{lunghi@mail.desy.de}}
\author{Antonio Masiero\\
SISSA -- ISAS and INFN, Sezione di Trieste, I-34013, Trieste, Italy\\
        E-mail: \email{masiero@he.sissa.it}}
\author{Werner Porod\\
        Institut f\"ur Theoretische Physik, 
  Universit\"at Z\"urich, CH-8057, Z\"urich, Switzerland \\
        E-mail: \email{porod@physik.unizh.ch}} 
\author{Hanns Stremnitzer\\
        Institut f\"ur Theoretische Physik, 
  Universit\"at Wien, A-1090, Vienna, Austria \\
        E-mail: \email{strem@ap.univie.ac.at}}
\author{Oscar Vives\\
Department of Physics, Theoretical Physics, U. of Oxford, Oxford
OX1 3NP, UK\\
       E-mail: \email{vives@thphys.ox.ac.uk}}

\abstract{We study FCNC and $CP$ violating processes in the MSSM
without a new flavour structure (flavour blind MSSM). The low energy
parameters are determined by the running of the soft breaking terms
from the GUT scale with SUSY phases consistent with the EDM
constraints. We find that the $CP$ asymmetry in $b\to s \gamma$ can
reach large values potentially measurable at B factories, especially
in the low $BR (b \to s \gamma)$ region. We analyze the SUSY
contributions to the anomalous magnetic 
moment of the muon pointing out its impact on the $b\to s\gamma$ $CP$
asymmetry and on the SUSY spectrum.}

\begin{document}
\newpage
In this contribution we study in a systematic way the
restrictions on the supersymmetry (SUSY) parameters and complex
phases which can be derived from the 
experimental information on flavour changing neutral currents
(FCNC) and on $CP$ violation. We choose the
Minimal Supersymmetric Standard Model (MSSM) as our theoretical
framework and focus on a class of SUSY models that we
call {\it flavour blind MSSM}. With this term we refer to a 
model where the soft breaking terms at the grand unification
(GUT) scale do not 
introduce any new flavour structure beyond the usual Yukawa
matrices. In this restricted class of models, the number of
parameters is largely reduced and it is therefore possible to
perform a complete phenomenological analysis. Here we present
the main results of our analysis. For details we refer
to~\cite{flabl}. 

We consider two cases, which are specified by
the structure of the soft breaking parameters at the GUT
scale. The
first case is the simplest version of the constrained 
MSSM, where we take the following independent parameters:\\
(I) $M_{1/2}$, $M_0^2$, $|A_0|$, $\phi_{A_0}$,
$\phi_{\mu}$, and $\tan\beta$,\\
which are the universal gaugino and scalar
mass parameter, the absolute value and the phase of the
trilinear scalar coupling, the phase of $\mu$, and the ratio of
the vacuum expectation values of the Higgs fields. The second
case refers to the SUSY $SU(5)$ model, where we take 
the following set of parameters:\\
(II) $M_{1/2}$, $M_{5}^2$, $M_{10}^2$, $M_{H_1}^2$,
$M_{H_2}^2$, $|A_u|$, $|A_d|$, $\phi_{A_u}$, $\phi_{A_d}$,
$\phi_{\mu}$, $\tan\beta$, \\ 
where now  $M_{5}^2$, $M_{10}^2$, $M_{H_1}^2$ and 
$M_{H_2}^2$ are the scalar mass parameters of the $\bar 5$ and
$10$ sfermions and the two Higgs doublets, and $A_u$ and $A_d$
are the trilinear scalar couplings of the $10$ and $\bar 5$.

With the SUSY parameters defined at the GUT scale we determine
the soft SUSY breaking parameters at the weak scale by evolving
them down with the renormalization group equations (RGE). In our
analysis, we have used two--loop RGEs as given
in~\cite{martin:1994zk} 
and one--loop masses as given in~\cite{pierce}. We fix
$|\mu|^2$ by demanding radiative breaking of 
the electroweak $SU(2)_L \times U(1)$ symmetry. At the weak
scale we 
impose the constraints from direct SUSY and Higgs particle
searches~\cite{PDG} and from the 
$\rho$--parameter, as well as the requirements of colour and
electric charge conservation and the lightest SUSY
particle (LSP) to be neutral. With these sets of soft SUSY
parameters we 
calculate the electric dipole moment (EDM) of the electron and
the branching ratio of 
$b \rightarrow s \gamma$ and compare them with the experimental
data, 
$|d^e| \leq 4.0 \times 10^{-27} \ e \ \mbox{cm}$ and 
$2 \times 10^{-4} \leq BR(b \to s \gamma) \leq
4.5 \times 10^{-4}$\cite{CLEO}.
The sets in agreement with the experimental constraints 
are used to calculate the $CP$ asymmetry of 
$b \rightarrow s \gamma$ and the SUSY contributions to
$\epsilon_K$, $\Delta M_{B_d}$ and $\Delta M_{B_s}$.
We also calculate the SUSY contributions
to the muon anomalous magnetic moment $a_{\mu^+}$ in order to
quantify the 
effect of the recent experimental data on the observables we are
interested in. We assume that the SUSY contribution accounts for
the difference between the experimental world average
for this quantity~\cite{brook} and
the corresponding SM prediction~\cite{marciano1},
$\delta a_{\mu^+} = +43(16) \times 10^{-10}$.

We first study the restrictions on the phases from the electron
EDM constraint. Due to cancellations between different
contributions these restrictions are less stringent and are very
different from the case that the EDM constraint is required for
each contribution separately 
(see~\cite{edm:we} and references therein). 
In our numerical calculations we scan the scalar and
gaugino masses at $M_{GUT}$ in the range 
$100\ \mbox{GeV} < M_i < 1\ \mbox{TeV}$, the trilinear terms 
$0 < |A_d|^2 < M^2_{10} + M^2_5 + M^2_{H_1}$, 
$0 < |A_u|^2 < 2 \, M^2_{10} + M^2_{H_2}$, taking their phases 
arbitrary, and $4 < \tan \beta < 50$.
We find that the EDM constraint implies correlations
between $\phi_{\mu}$ and $\phi_{A_0}$ and between
$\phi_{\mu}$ and $M_0^2$. While it is possible to find any value
for $\phi_{A_0}$, the values of $\phi_{\mu}$ are more
constrained: $\phi_\mu \lsim 0.1$ for lower values of $M_0^2$,
whereas values up to $\phi_{\mu} = 0.4$ are allowed for higher
$M_0^2$. 

Next we discuss the main features of the SUSY particle spectrum 
relevant for $CP$ violating and FCNC observables. 
In Fig.~\ref{fig:mchmstgm2} we show the scatter plot of
the lightest chargino versus the lightest stop masses for
the SUSY $SU(5)$ model (set (II)). All points fulfill the EDM
and $BR(b \to s \gamma)$ constraints, whereas the black dots are
the points of the parameter space that also fulfill the
$\delta a_{\mu^+}$ constraint.
We confirm the presence of an upper bound on the
chargino mass of about $700\ \mbox{GeV}$ for very large 
$\tan\beta$ (of order 50) 
\cite{gm2pap}, and lower for smaller values of $\tan \beta$,
which is a consequence of the $\delta a_{\mu^+}$ constraint. It 
can be shown that this bound is essentially due to our
assumption of 
gaugino mass unification. We would like to stress, however, that
in any RGE evolved MSSM with gaugino mass unification the
$a_{\mu^+}$ constraint has important consequences on the
complete MSSM spectrum. As can be seen in
Fig.~\ref{fig:mchmstgm2}, there is a strong correlation among
the stop and chargino masses due to the dominant gaugino RGE
effects. Then, without any further restriction on the
SUSY parameter space, this correlation
implies the presence of an upper bound on the light 
stop mass of $m_{\tilde{t}} \leq 1500\ \mbox{GeV}$.

An especially interesting observable is the $CP$ asymmetry in the
partial width,\\
$$A_{CP}^{b\rightarrow s\gamma} =
{BR(\bar B\rightarrow X_s \gamma) - BR(B\rightarrow X_s \gamma) \over
 BR(\bar B\rightarrow X_s \gamma) + BR(B\rightarrow X_s \gamma)}.$$\\
The SM prediction for this asymmetry is exceedingly small,
therefore, it is sensitive to
the new SUSY phases. As $\phi_\mu$ and
$\phi_A$ are associated with chirality changing
operators, we can expect large effects in $b \to s \gamma$
\cite{flavor}. 
In Fig.~\ref{fig:acpamu} we show 
the correlation among $\delta a_\mu$ and $\mbox{BR}( b \to s
\gamma)$. All points shown fulfill the EDM constraint, whereas
the black dots represent the points of the parameter space that
also reproduce the measured anomalous magnetic moment.
In this plot we can see that the $\delta a_\mu$ constraint cuts
out the region of high branching ratios $BR(b \to s \gamma)$. In
our flavour 
blind scenario the main contributions to the $b \to s \gamma$
decay are due to charged Higgs, light stop and light chargino
loops. The $BR(b \to s \gamma)$ constraint and the 
$\delta a_\mu$ constraint favour $Re(\mu) > 0$. Imposing the EDM
constraint reduces the $CP$ asymmetry to less than $1\%$ in the
region of large branching ratio $BR(b \to s \gamma)$. However,
for small $BR(b \to s \gamma)$ the $CP$ asymmetry can go up to
$5\%$, making it potentially measurable at B factories.

Finally we calculated the SUSY contributions to the
$CP$ violating quantities $\epsilon_K$, $\Delta M_{B_d}$,
$\Delta M_{B_s}$ and studied the modifications of the unitarity
triangle. We obtain only small deviations from the SM predictions.
This result is different from those obtained in the context of
Minimal Flavour Violation MSSM or in the MSSM with
non--universal parameters ~\cite{mfv,luvi}

{\sc Acknowledgements:} This work was supported
by the `Fonds zur F\"orderung der wissenschaftlichen Forschung' of
Austria, FWF Project No.  P13139-PHY, by the EU TMR Network Contracts
No. HPRN-CT-2000-00148, HPRN-CT-2000-00149 and HPRN-CT-2000-00152 and
by Cooperazione scientifica e tecnologica Italia-Austria 1999-2000,
Project No. 2; W.P. was supported by the Spanish Ministry of Education
and Culture under the contract SB97-BU0475382 and by DGICYT grant
PB98-0693; T.G. acknowledges financial support from the DOE grant
DE-FG02-96ER40967; E.L. and O.V. thank SISSA for support during the
first stages of this work. E.L. was supported by the Alexander Von
Humboldt Foundation; O.V. acknowledges financial support from a Marie
Curie E.C. grant (HPMF-CT-2000-00457) and partial support from Spanish
CICYT AEN-99/0692.

\FIGURE{\epsfig{file=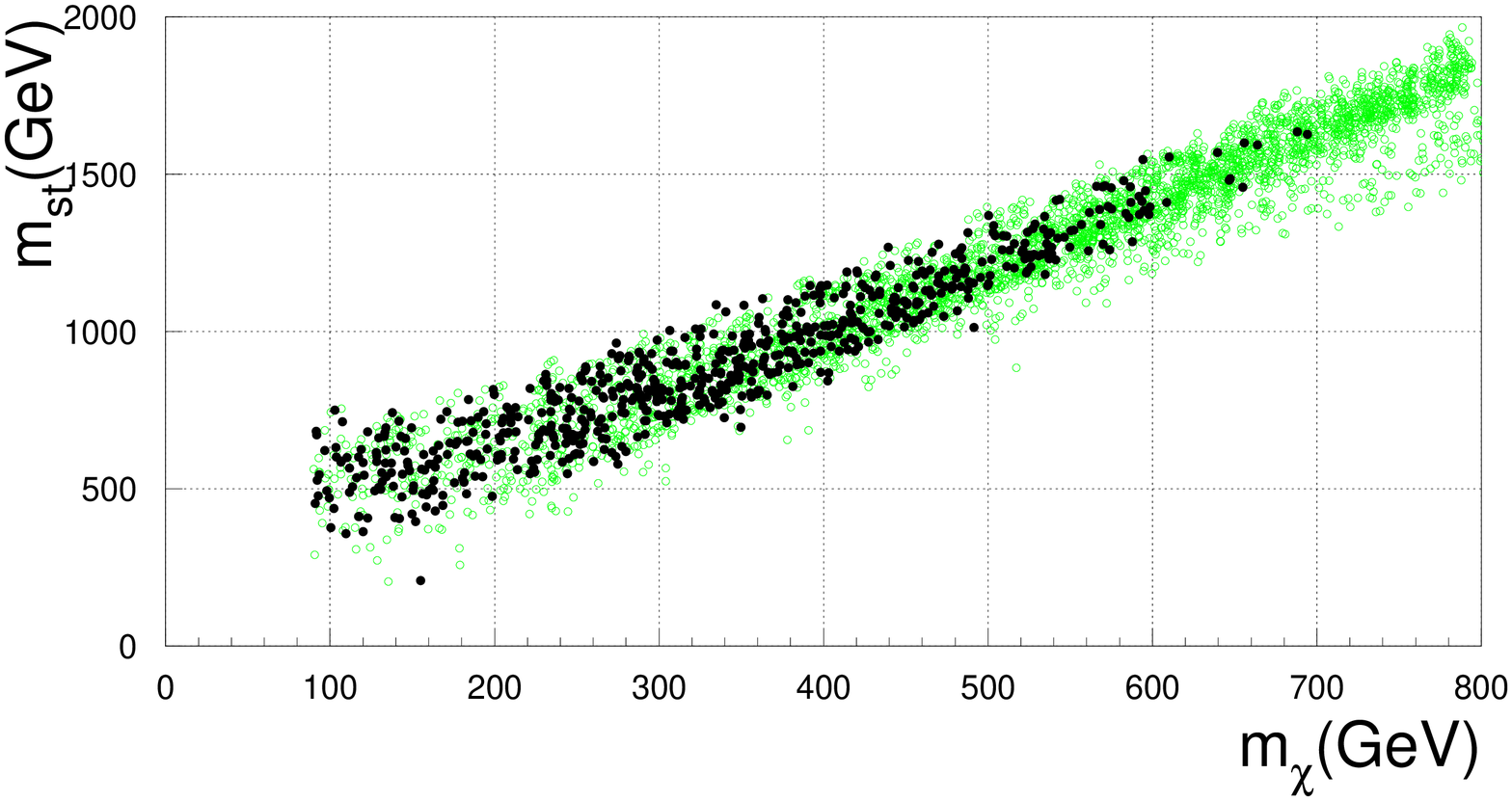,width=0.8 \linewidth} 
\caption{
Chargino--stop mass correlations in the $SU(5)$ inspired
MSSM with $b \to s \gamma$ constraint satisfied. The black dots
satisfy also the $a_{\mu^+}$ constaint.\label{fig:mchmstgm2}}}

\FIGURE{\epsfig{file=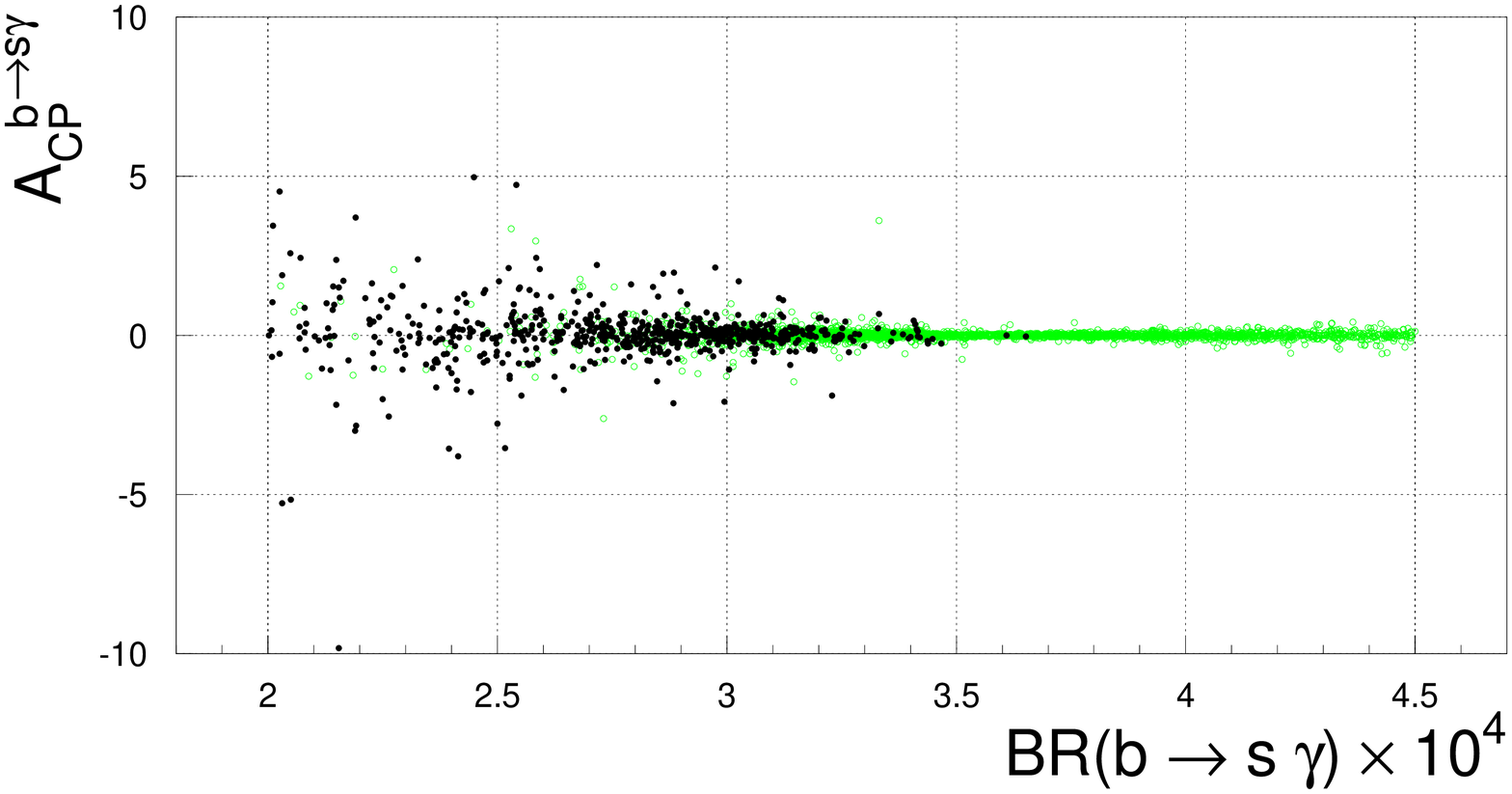,width=0.8 \linewidth}
\caption{Plot of the $CP$ asymmetry versus the branching ratio of
$b\rightarrow s \gamma$. We allow only for points whose phases satisfy
the EDM's constraints.  The black dots satisfy the $a_{\mu^+}$
constraint.}
\label{fig:acpamu}}


%

\end{document}